\documentclass[epj%,referee
]{svjour}
\usepackage{graphicx}
\usepackage{amssymb}
\begin{document}
\titlerunning{Calculation of Critical Perturbation Amplitudes and Critical Densities of Traffic Flows}
\title{Analytical Calculation of Critical Perturbation Amplitudes and Critical Densities by Non-Linear Stability Analysis of a Simple Traffic Flow Model}
\author{Dirk Helbing and Mehdi Moussaid% etc
}                     % Do not remove
\institute{ETH Zurich, UNO D11, Universit\"atstr. 41, 8092 Zurich, Switzerland}
\date{Received: date / Revised version: date}
\abstract{Driven many-particle systems with nonlinear interactions are known to often display multi-stability,
i.e. depending on the respective initial condition, there may be different outcomes. Here, we study this phenomenon for traffic models, some of which show stable and linearly unstable density regimes, but areas of metastability in between. 
In these areas, perturbations larger than a certain critical amplitude will cause a lasting breakdown of traffic, while smaller 
ones will fade away. While there are common  methods to study linear instability, non-linear instability had to be studied {\it numerically} in the past. Here, we present an {\it analytical} study for the optimal velocity model with a stepwise specification of the optimal velocity function and a simple kind of perturbation. Despite various approximations, the analytical results are shown to reproduce numerical results very well.
 % end of PACS codes
} %end of abstract
\maketitle
\section{Introduction}

While the field of traffic modeling and traffic simulation has a long history \cite{Gazis}, in the 90ies \cite{NaSch,BML,KKfirst}, it has also become interesting for a large community of physicists. Since then, various physical methods have been applied, ranging from fluid-dynamic and gas-kinetic approaches \cite{Flui,GaFl,Ga} over many-particle models \cite{Review} up to cellular automata \cite{NagBook}. An overview of the respective literature has been given by various reviews \cite{Review,Schadschneider,Nagatani}.
\par
Besides computer-based studies, physicists have particularly contributed with systematic (e.g. gas-kinetic) derivations \cite{Ga} and analytical investigations \cite{analy} (see Ref. \cite{Review} for an overview). This includes the study of instabilties leading to a breakdown of free traffic flow \cite{Heretal,Band,Mitarai,Wilson1,Wilson2,Wilson3}, which has been studied experimentally only recently \cite{Sugi}. (Of course, empirical studies \cite{KBook,TraSci} have been already carried out for a longer time).
\par
While there are also models, where the instability of traffic flow depends on noise \cite{NaSch}, we will focus here on models with a deterministic instability mechanism. Many traffic models become unstable in a certain density range, because of delays in the adaptation to changing traffic conditions. To avoid accidents, these delays are compensated for by over-reactions, which can increase a small initial perturbation and finally cause a breakdown of traffic flow. This kind of dynamics usually occurs in the medium density range, where the change of the ``desired'', ``optimal'', or ``equilibrium'' velocity with a change in the distance or density is larger than a certain instability threshold \cite{EP}. 
\par
To characterize the behavior of traffic flows more systematically, Kerner and Konh\"auser  
have applied the notion of critical densities and critical perturbation amplitudes. Primarily based on numerical studies  \cite{KKPre94}, they found the following for a macroscopic, Navier-Stokes-like traffic model \cite{Review}: Altogether, there are four critical densities $\rho_{{\rm c}k}$ with $k\in \{1,2,3,4\}$. Below some density $\rho_{\rm c1} > 0$, any kind of disturbance eventually disappears.
Between the densities $\rho_{\rm c1}$ and $\rho_{\rm c2}$, 
one wide traffic jam builds up, given a large enough perturbation. A series of traffic jams may
appear in a density range between $\rho_{\rm c2}$ and some density
$\rho_{\rm c3}$. Finally, a so-called ``anticluster''  can be triggered \cite{KKPre94}, if the average density $\overline{\rho}$ is
between $\rho_{\rm c3}$ and some critical density $\rho_{\rm c4}$, while any disturbance disappears 
in stable traffic above $\rho_{\rm c4}$. 
Similar observations have been made for other macroscopic traffic models \cite{McM}, but also microscopic ones of the car-following or cellular automata type \cite{BandoKerner,McM,Cellu}. The critical densities $\rho_{{\rm c}k}$ depend mainly on the choice of the model parameters, in particular the relaxation time and the velocity-distance or velocity-density relation. 
\par\begin{figure}[htbp]
\begin{center}
\end{center}
\caption{\small(Color Online) Formation of a traffic jam in case of overcritical perturbations (blue lines) and relaxation to free traffic flow in case of undercritical perturbations (red lines) for a low vehicle density. The plot shows 10 trajectories representing each 10th vehicle. Altogether, the simulation was performed with N=100 vehicles on a circular road of length $L=N/\rho$. Time is measured in units of the relaxation time $\tau$, vehicle locations in units of the safe distance $d_0$. The slope of the lines corresponds to the vehicle speeds in units of $d_0/\tau$, and their vertical separation reflects the vehicle distances in units of $d_0$. Therefore, the blue lines correspond to stop-and-go traffic, where the vehicle distances in the stop regime are small, i.e. vehicle density is large. The red trajectories show that the same number of vehicles can constantly move at the free speed $v^0$, if the initial perturbation is small enough.}
\label{traje1}
\end{figure}
If the average density $\overline{\rho}$ falls into 
the density ranges $[\rho_{\rm c1}, \rho_{\rm c2}]$ or $[\rho_{\rm c3},\rho_{\rm c4}]$,
traffic flow is predicted to be {\it metastable}, i.e. 
characterized by a {\em critical amplitude} $\Delta \rho_{\rm cr}(\overline{\rho})$
for the formation of traffic jams. This amplitude is zero for $\overline{\rho} =
\rho_{\rm c2}$ and $\overline{\rho} =\rho_{\rm c3}$, i.e. at the boundaries of the linearly {\it unstable} regime,
while the critical amplitude grows towards the boundaries $\rho_{\rm c1}$ and $\rho_{\rm c4}$ of {\it stable} traffic and does not exist beyond these values. Perturbations with subcritical amplitudes $\Delta \rho < \Delta \rho_{\rm cr}(\overline{\rho})$ 
are eventually damped out (analogous to the stable density ranges),
while perturbations with supercritical amplitudes $\Delta \rho > \Delta \rho_{\rm cr}(\overline{\rho})$ 
grow and form traffic jams (similar to the linearly unstable density
ranges).  The situation in metastable traffic is, therefore, similar to supersaturated vapor \cite{KKPre94}, where an overcritical nucleus is required for condensation {\em (``nucleation effect'')}.
\par\begin{figure}[htbp]
\begin{center}
\end{center}
\caption{\small (Color Online) Formation of stop-and-go traffic from an initial perturbation in linearly unstable traffic flow. For details see the main text and the caption of Fig. \ref{traje1}.}
\label{traje2}
\end{figure}
In order to gain analytical insights into the mechanisms of nonlinear instability, we will proceed as follows:
In Sec. \ref{ovm}, we will introduce the optimal velocity model we will work with. Afterwards, in Sec. \ref{critdef}, we will discuss how perturbation sizes can be measured and what are {\it necessary} preconditions for the existence of traffic jams. We will then determine {\it sufficient} conditions by calculating critical densities and critical perturbation amplitudes for a simple kind of perturbations: In Sec. \ref{large}, all vehicles on a circular road but one are assumed to be in a big traffic jam, while in Sec. \ref{small}, all vehicles but one are assumed to experience free flow conditions. Finally, Sec. \ref{sumi} will summarize our results and present an outlook. Supplementary, and for comparison with our results, the Appendix gives a short derivation of the characteristic constants of the car-following model studied in this paper.
\begin{figure}[htbp]
\begin{center}
\end{center}
\caption{\small (Color Online) Formation of a traffic jam in case of an overcritical perturbation (blue lines) and drop of the vehicle flow to zero in case of undercritical perturbations (blue lines) for a large vehicle density. For details see the main text and the caption of Fig. \ref{traje1}.}
\label{traje3}
\end{figure}

\section{Introduction of the Applied Car-Following Model}\label{ovm}

In the past, a large number of papers has addressed the instability of traffic models in an analytical way,
and our discussion naturally needs to restrict itself to a few of them. For example, Kerner {\it et al.} \cite{Asymptotic} have presented an asymptotic theory of traffic jams for a Navier-Stokes-like, macrocopic traffic model, while instability analyses for car-following models were carried out by Herman {\it et al.} \cite{Heretal}, Bando {\it et al.} \cite{Band}, and others (see citations in Ref. \cite{Review}). 
Some recent analyses of traffic instabilities are by Mitarai and Nakanishi \cite{Mitarai} and by Wilson \cite{Wilson1,Wilson2,Wilson3}.
The most relevant references on which this paper builds on are Refs. \cite{Exact,Cellu}. For a car-following model and a related cellular automaton, respectively, these have presented analytical calculations of the characteristic constants of traffic flows such as the dissolution speed $c$ of traffic congestion (which agrees with the propagation speed of traffic jams), the outflow $Q_{\rm out}$ from congested traffic, or the vehicle density $\rho_{\rm jam}$ in congested traffic for a car-following model and a related cellular automaton, respectively. 
\par
The success in determining the self-organized constants of traffic flow analytically is based on boiling down vehicle dynamics to its essence. That is, we will work with a very simple model (a ``toy model''), that is not aiming to be fully realistic, but instead at performing analytical calculations and deriving stylized facts that would not be analytically accessible for a more realistic model with a reasonable amount of effort. The model we will focus on is the car-following model by Bando {\it et al.} \cite{Band} with a stepwise optimal velocity function. It is defined by 
the equation of motion $dx_j/dt = v_j(t)$ relating the change of location $x_j(t)$ of a vehicle $j$ in time $t$ with its speed $v_j(t)$ and the acceleration equation
\begin{equation}
\frac{dv_j}{dt} = \frac{v_{\rm o}\big(d_j(t)\big) - v_j(t)}{\tau} \, . \label{cafo}
\end{equation}
The parameter $\tau >0$ corresponds to a small ``relaxation time''. The ``optimal velocity function'' $v_{\rm o}(d)$ depends on the vehicle distance $d_j(t) =x_{j-1}(t) - x_j(t)$ of a vehicle to its predecessor (``leader'') $j-1$, and it is specified here as
\begin{equation}
v_{\rm o}(d) = \left\{
\begin{array}{ll}
v^0 & \mbox{if } d> d_0, \\
0 & \mbox{otherwise.}
\end{array}\right.
\label{smpl}
\end{equation}
$d_0\gtrsim v^0\tau$ is a safe vehicle distance and $v^0$ the free speed (or speed limit) of the vehicles, which are assumed to be identical, here. Defining the local vehicle density $\rho$ as the inverse of the vehicle distance and the stationary and homogeneous equilibrium flow as $Q_{\rm e}(\rho) = \rho v_{\rm o}(1/\rho)$, we find
\begin{equation}
Q_{\rm e}(\rho) = 
\left\{\begin{array}{ll}
\rho v^0 & \mbox{if } \rho <  1/d_0, \\
0 & \mbox{otherwise.}
\end{array}\right.
\end{equation}
It is obvious that this flow-density relationship (``fundamental diagram'') is not realistic. However, when the traffic flow is unstable with respect to perturbations in the flow, a much more realistic, self-organized flow-density relation results, namely the so-called  ``jam line''  \cite{Asymptotic} 
\begin{equation}
J(\rho)  = \frac{1}{\rho_{\rm jam}T} (\rho_{\rm jam} - \rho ) 
\end{equation}
(see Fig. \ref{instabdiag}).
The jam density $\rho_{\rm jam}$ and the delay $T$ between the acceleration of successive vehicles at the jam front can even be calculated (see Ref. \cite{Exact} and the Appendix of this paper). The corresponding implicit relations are 
\begin{equation}
 \frac{1}{\rho_{\rm jam}} = d_0 - v^0 \tau \big( 1 - \mbox{e}^{-T/\tau}\big) 
\label{jamdef}
\end{equation}
and 
\begin{equation}
 T=  2 \tau\Big( 1 - \mbox{e}^{-T/\tau}\Big) \, .
 \label{Tedef}
\end{equation}
It is also worth stating that the characteristic outflow 
\begin{equation}
Q_{\rm out} = \rho_{\rm out} v^0 = \frac{1}{d_0/v^0 + T/2} 
\end{equation}
from congested traffic  is smaller than the maximum flow 
\begin{equation}
Q_{\rm max} = \frac{v^0}{d_0} \, , 
\end{equation}
that is, there must be a density region in which two different flows are possible. Hence, the actually assumed density value depends on the history. This is called a {\it hysteresis effect}. Once a traffic jam forms, there is an effective {\em capacity drop} of size
\begin{eqnarray}
\Delta Q_{\rm drop} &=& Q_{\rm max} - Q_{\rm out} =
\frac{v^0}{d_0} - \frac{1}{d_0/v^0 + T/2}  \nonumber \\
&=& \frac{v^0T/2}{d_0(d_0/v^0 + T/2)} > 0   \label{DroP}
\end{eqnarray}
(see Fig. \ref{instabdiag}).
\begin{figure}[htbp]
\begin{center}
\end{center}
\caption{\small Schematic illustration of the traffic flow as
    a function of the vehicle density $\rho$.  The density ranges between $\rho_{\rm c1}$ and $\rho_{\rm c2}$
    and between $\rho_{c3}$ and $\rho_{\rm c4}$ correspond to metastable traffic flow (see main text). The quantities
    $Q_{{\rm c}k}=Q(\rho_{{\rm c}k})$ denote the flows belonging to
    the critical densities $\rho_{{\rm c}k}$, which will be analytically calculated in this paper. The thin solid line corresponds to the jam line
    $J(\rho)$. Its intersection point with the free branch of the fundamental diagram defines the value of the characteristic
    outflow $Q_{\rm out}$ and the density $\rho_{\rm out}$, while its intersection point with $Q=0$ defines the jam density $\rho_{\rm jam}$.
    The fundamental diagram, i.e. the flow-density relation  $Q_{\rm e}(\rho)$ in the stationary and homogeneous case, is represented by thick solid lines. The difference between the maximum flow $Q_{\rm max} = \max_\rho Q_{\rm e}(\rho)$ and the characteristic outflow $Q_{\rm out}$ is called the capacity drop $\Delta Q_{\rm drop}$.}
\label{instabdiag}
\end{figure}
 
\section{Definition of Critical Amplitudes and Critical Densities}\label{critdef}

It is very encouraging that characteristic constants of traffic flows such as the jam density $\rho_{\rm jam}$, the outflow $Q_{\rm out}$ from congested traffic, and the propagation speed $c = -1/(\rho_{\rm jam}T)$ of jam fronts can be analytically derived from the optimal velocity model.
This just requires a simple enough specification (\ref{smpl}) of the function $v_{\rm o}(d)$, see Ref. \cite{Exact} and the Appendix of this paper.
Therefore, we will try in the following to derive the critical densities and critical amplitudes for this model as well. For this, let us study a circular one-lane system of length $L$ with $N$ vehicles and an average vehicle density $\overline{\rho} = N/L$. In order to avoid finite size effects, we will assume a large system with many vehicles. According to Eqs. (\ref{cafo}) and (\ref{smpl}), if all vehicles have a distance greater than $d_0$ to their leader, all of them will accelerate and reach the maximum speed $v^0$, while we expect congested, standing traffic, if all vehicles have a distance smaller than $d_0$, since this forces them to decelerate. 
\par
Therefore, an interesting dynamics will only occur if some vehicles have distances larger than $d_0$, while others have distances smaller than $d_0$. In the following, we will focus on this case. One first observation is that {\em linearly} unstable traffic should be possible only for $\overline{\rho} = N/L = 1/d_0$, so that, for the stepwise specification (\ref{smpl}) of the optimal velocity function $v_{\rm o}(d)$, we have the special case
\begin{equation}
 \rho_{\rm c2} = \rho_{\rm c3} = \frac{1}{d_0} \, .
\end{equation}
But what is the value of the critical density $\rho_{\rm c1}$ at which non-linear instability starts to be possible,
and is there a critical density $\rho_{\rm c4}$, beyond which even arbitrarily large perturbations will fade away?
And can we approximately determine the critical amplitudes $\Delta \rho_{\rm cr}(\overline{\rho})$?
\par
Let us in the following focus on a special kind of perturbations (see Figs. \ref{First} and \ref{Second}): We will assume that $(N-1)$ vehicles have an identical distance $d_{(0)}$ to their respective predecessor, while the distance $d_1(0)$ of the first vehicle to its predecessor is 
\begin{equation}
d_1(0) = L - (N-1)d_{(0)} \, . 
\end{equation}
Accordingly, we define the perturbation size as
\begin{eqnarray}
 \Delta \rho &=& \left| \frac{1}{d_{(0)}} - \frac{1}{d_1(0)} \right| =  \left| \frac{1}{d_{(0)}} - \frac{1}{L - (N-1)d_{(0)}} \right| \nonumber \\
 &=& \left| \frac{L - Nd_{(0)}} {d_{(0)} [L - (N-1)d_{(0)}]} \right| \, . 
 \label{criamp}
\end{eqnarray}
In order to avoid finite-size effects, we will assume a very large system of length $L = N/\overline{\rho}$ with $N\gg 1$ vehicles. Then, we have 
\begin{equation}
d_{(0)} = \frac{L - d_1(0)}{N-1} \rightarrow \frac{L}{N} = \frac{1}{\overline{\rho}} \, ,
\label{densconv}
\end{equation}
which implies
\begin{equation}
\Delta \rho \approx \left| \overline{\rho} - \frac{1}{d_1(0)} \right| \, . 
\label{whichimplies}
\end{equation}
The critical {\it amplitudes} $\Delta \rho_{\rm cr}$ are basically defined by not changing in time (neither growing nor shrinking), i.e. by {\it marginal stability}. Moreover, the critical {\it density} $\rho_{\rm c1}$ is characterized by the fact that the critical amplitude ceases to exist for lower densities, and the same applies for densities greater than the critical density $\rho_{\rm c4}$. 
\par
Remember that the propagation of perturbations requires at least one vehicle distance (either $d_{(0)}\approx 1/\overline{\rho}$ or $d_1(0)$) to be above and another one below $d_0$. This allows us to estimate a lower bound for the critical perturbation amplitudes $\Delta \rho_{\rm cr}(\overline{\rho})$. Two cases my be distinguished:
\begin{enumerate}
\item If the density $\overline{\rho}\approx 1/d_{(0)}$ is greater than $1/d_0$, we must have $d_1(0) > d_0$, which implies $[\overline{\rho} - 1/d_1(0)] > \overline{\rho} - 1/d_0$.
\item If the density $\overline{\rho}\approx 1/d_{(0)}$ is smaller than $1/d_0$, we must have $d_1(0) < d_0$, which implies $[1/d_1(0)-\overline{\rho}] > 1/d_0 - \overline{\rho}$.
\end{enumerate} 
Consequently, the critical amplitude can become zero only for $\overline{\rho} = 1/d_0 = \rho_{\rm c2} = \rho_{\rm c3}$, 
and altogether we must have  
\begin{equation}
 \Delta \rho_{\rm cr}(\overline{\rho}) \ge  \left| \overline{\rho} - \frac{1}{d_0} \right| \, . 
\label{nodiverge}
\end{equation}
However, this is only a {\it necessary} condition, while we will determine {\it sufficient} conditions for the existence of stop-and-go waves in the following sections. In our further analysis, we will treat the case of large densities separately from the case of small densities.

\section{The Case of Large Densities} \label{large}

The situation for densities $\overline{\rho} > 1/d_0$ is illustrated in Fig. \ref{First}. We assume that all vehicles $j$ start at time $t=0$ with speed $v_j(0) = 0$. Moreover, all vehicles but one are assumed to have the initial distance $d_{(0)} < d_0$, while the remaining vehicle $j=1$ has the distance $d_1(0) = L - (N-1)d_{(0)} >d_0$, where $L$ is the length of the assumed circular road. It will turn out that the simplicity of this initial perturbation is the reason for the feasibility of our calculations. Without loss of generality, vehicle $j=1$ starts to accelerate at $x=0$ and $t=0$. 
\par
We will now have to identify the possible reason for a disappearance of the initial perturbation in the course of time.
In the case of large densities $\overline{\rho} > \rho_{\rm c3} = 1/d_0$, {\it an initial perturbation will fade away if the maximum vehicle distance does not allow to reach a sufficiently high speed in the acceleration process to reach a vehicle distance equal to or smaller than $d_{(0)}$ after its successive braking maneuver.} While the speed in the vehicle queue is approximately zero (despite for, maybe, the last few vehicles in the queue which may still decelerate), the maximum speed $v_{(0)}$ is reached at the time $t_{(0)}$ when the vehicle under consideration starts to decelerate. Focussing on vehicle $j=1$, we find
\begin{equation}
 v_{(0)} = v_1(t_{(0)}) = v^0 \big( 1 - \mbox{e}^{-t_{(0)}/\tau} \big) \, , 
 \label{ve0}
\end{equation}
because Eqs. (\ref{cafo}) and (\ref{smpl}) imply $v_1(t) = v^0 (1 - \mbox{e}^{-t/\tau})$, which can be easily checked by differentiation with respect to $t$.
\par\begin{figure}[htbp]
\begin{center}
\end{center}
\caption[]{Illustration of the initial distribution of vehicles on a ring road assumed in the high-density case ($\overline{\rho}>1/d_0$). Vehicles $j\in \{2,\dots,N\}$ are standing in a traffic jam with speed $v_j = 0$ and a distance $d_{(0)}<d_0$ to the predecessor. Vehicle $j=1$ starts with zero velocity as well, but can accelerate, if its distance $d_1(0)$ to the next vehicle is greater than $d_0$.}
\label{First}
\end{figure}
As the first vehicle's position at time $t=0$ is assumed to be $x_1(0)=0$, the time-dependent location of vehicle 1 is given by
\begin{eqnarray}
 x_1(t) &=& \int\limits_0^t dt' v_1(t') = \int\limits_0^t dt' \; ( v^0 - v^0\mbox{e}^{-t'/\tau}) \nonumber \\
 &=& v^0t + v^0 \tau \Big(\mbox{e}^{-t/\tau} - 1 \Big) \, .
 \label{courseq}
\end{eqnarray} 
The time $t_{(0)}$ at which vehicle $j=1$ starts to decelerate is the time $t$ at which its distance to vehicle $j=N$, i.e. the last vehicle in the queue, becomes $d_0$. Since vehicle $j=N$ is located at $x_N=L - (N-1)d_{(0)}$, this implies
\begin{equation}
 L - (N-1)d_{(0)} - x_1(t_{(0)}) = d_0
\end{equation}
or
\begin{equation}
L - (N-1)d_{(0)} - d_0 = v^0\Big(t_{(0)} - \tau + \tau \mbox{e}^{-t_{(0)}/\tau}\Big) \, ,
\label{Lminus}
\end{equation}
which is an implicit equation determining the acceleration time period $t_{(0)}$. At the end of its deceleration process, the previously first vehicle of the queue, which has then joined the end of the queue, will have a distance $d_{(1)}$, which depends on the maximum speed $v_{(0)}$. Since the deceleration process according to Eqs. (\ref{cafo}) and (\ref{smpl}) obeys an exponential velocity decay
\begin{equation}
v_1(t) = v_{(0)} \mbox{e}^{-(t-t_{(0)})/\tau} \, ,
\end{equation}
the resulting minimum distance can be determined as
\begin{equation}
 d_{(1)} = d_0 - \int\limits_{t_{(0)}}^\infty dt \; v_{(0)} \mbox{e}^{-(t-t_{(0)})/\tau}
  = d_0 - v_{(0)} \tau \, . 
  \label{de1}
\end{equation}
The other vehicles in the queue are expected to have the same distance to their respective predecessor after one cycle of acceleration and deceleration. That is, while their distance was $d_{(0)}$ in the beginning, their distance after one cylce will be $d_{(1)}$, after two cycles it will be $d_{(2)}$, and so on.
The iterative equations to determine the decisive quantities can be derived analogously to Eqs. (\ref{de1}) and (\ref{ve0}):
\begin{equation}
 d_{(n+1)} = d_0 - v_{(n)} \tau = d_0 - v^0 \tau  \big( 1 - \mbox{e}^{-t_{(n)}/\tau} \big) \, ,
 \label{consider1}
\end{equation}
where $t_{(n)}$ as a function of $d_{(n)}$ is determined from
\begin{equation}
L - (N-1)d_{(n)} - d_0 = v^0\Big(t_{(n)} - \tau + \tau \mbox{e}^{-t_{(n)}/\tau}\Big)  
\label{consider2}
\end{equation}
similarly to Eq. (\ref{Lminus}).
Equation (\ref{consider2}) allows us to replace the right-hand side of Eq. (\ref{consider1}), 
which yields an equation for $d_{(n+1)}$ as a function of $d_{(n)}$:
\begin{equation}
 d_{(n+1)}  = L - (N-1)d_{(n)} - v^0 t_{(n)}(d_{(n)}) \, . \label{consider3} 
\end{equation}
It is, therefore, interesting to ask,
whether the series $d_{(n)}$ converges and, if yes, to what value. If the values $d_{(n)}$ stay the same for different values of $n$, the initial perturbation is stable over time. If the distances go down, then the perturbation grows. 
However, if the values of $d_{(n)}$ grow with $n$, the initial perturbation fades away (which is expected to happen, when the perturbation is too small). Therefore, {\it the critical amplitude is given by the condition $d_{(n+1)} = d_{(n)}$ of marginal stability}, which together with Eq. (\ref{consider3}) 
implies 
$d_{(n)}  = L - (N-1)d_{(n)} - v^0 t_{(n)}$
or
\begin{equation}
 v^0 t_{(n)} 
 = L - Nd_{(n)} \, .
 \label{consider4}
\end{equation}
Multiplying Eq. (\ref{consider1}) with $(N-1)$, 
we obtain for the marginally stable case $d_{(n+1)} = d_{(n)}$:
\begin{equation}
(N-1) d_{(n)} = (N-1) \Big[ d_0 -  v^0\tau \big( 1 - \mbox{e}^{-t_{(n)}/\tau} \big) \Big] \, .  
\end{equation}
Inserting Eq. (\ref{consider2}) finally yields
\begin{equation}
v^0 t_{(n)} =  L - Nd_0 + Nv^0\tau \big( 1 - \mbox{e}^{-t_{(n)}/\tau} \big)  \, . 
 \label{divine}
\end{equation}
Dividing Eq. (\ref{divine}) by $N$, considering $L/N = 1/\overline{\rho}$, and performing the limit $N\rightarrow \infty$ gives 
\begin{equation}
 v^0 \tau \big( 1 - \mbox{e}^{-t_{(n)}/\tau} \big) = d_0 - \frac{1}{\overline{\rho}} \, , 
 \end{equation}
i.e. one solution with a finite value of $t_{(n)}$:
\begin{equation}
t_{(n)} = -  \tau \ln \left( 1 -  \frac{d_0 - 1/\overline{\rho}}{\tau v^0} \right) 
 \approx \frac{1}{v^0} \left( d_0 - \frac{1}{\overline{\rho}} \right) \, . 
 \label{thesolution}
\end{equation}
Here, our approximation is based on the first-order Taylor expansion $\ln (1-x) \approx -x$.
Together with Eq. (\ref{consider4}), we have
\begin{equation}
d_0 - \frac{1}{\overline{\rho}} \approx  v^0 t_{(n)} 
 = L - Nd_{(n)} \, ,
\end{equation}
and considering $d_{(n+1)} = d_{(n)} = \dots = d_{(0)}$ finally gives
\begin{equation}
L - Nd_{(0)} \approx  d_0 - \frac{1}{\overline{\rho}} \, .
\end{equation}
This can now be inserted into Eq. (\ref{criamp}) to obtain the critical amplitudes:
\begin{eqnarray}
\Delta \rho_{\rm cr}(\overline{\rho})  &=& \left| \frac{L - Nd_{(0)}} {d_{(0)} [L - (N-1)d_{(0)}]} \right| \nonumber \\
 &\approx & \left| \frac{d_0 - 1/\overline{\rho}} {d_{(0)} ( d_0 - 1/\overline{\rho} +d_{(0)})} \right| \nonumber \\
 &\approx & \left| \frac{d_0 - 1/\overline{\rho}} {d_0/\overline{\rho}} \right| 
 = \left|  \overline{\rho} - \frac{1}{d_0} \right| \, , 
 \label{biig}
\end{eqnarray}
where we have applied $d_{(0)} \approx 1/\overline{\rho}$ according to Eq. (\ref{densconv}). 
\par
Equation (\ref{biig}) agrees well with our numerical findings (see Fig. \ref{numfind}). Obviously, the critical perturbation amplitude is zero for $\rho = 1/d_0 = \rho_{\rm c3}$, as it should. Moreover, for the specification (\ref{smpl}) of the optimal velocity function, the critical amplitude does not diverge at a finite density. Nevertheless, there is a critical density $\rho_{\rm c4}$, which is given by the fact that $d_{(n+1)} \ge d_0 - v^0\tau$ according to 
Eq. (\ref{consider1}). Considering $d_{(n+1)} = d_{(0)} \approx 1/\overline{\rho}$ in the marginally stable case, a critical amplitude
ceases to exist for densities larger than
\begin{equation}
\rho_{\rm c4} = \frac{1}{d_0 - v^0\tau } \, . \label{rc4}
\end{equation}
\begin{figure}[htbp]
\begin{center}
\end{center}
\caption[]{(Color Online) Results of computer simulations of the car-following model (\ref{cafo}) with the optimal velocity function (\ref{smpl}) for a time discretization $\delta t = 0.1$ and $v^0 = d_0/\tau$ (where $d_0=1$ and $\tau = 1$ has been assumed because of the possibility to scale space and time). The simulations were run for $N=100$ vehicles, with the initial conditions illustrated in Figs. \ref{First} and \ref{Second}. Given a certain average density $\overline{\rho}$, the length $L$ of the simulation stretch was chosen as $L=N/\overline{\rho}$, and the perturbation amplitude $\Delta \rho_{\rm cr}(\overline{\rho}) = | 1/d_{(0)} - 1/[L-(N-1)d_{(0)}]|$ determined the initial distance $d_{(0)}$ between $N-1$ of the vehicles. The bold 
dashed lines represent the necessary condition (\ref{nodiverge}) for the existence of stop-and-go waves, while the thin solid lines represent the approximate critical amplitudes calculated in this paper. The critical amplitudes according to our numerical simulations lie between the areas represented by different kinds of symbols reflecting different traffic states at $t=2000$:  green plus signs mean all vehicles move at the desired speed $v^0$, blue circles mean all vehicles have stopped moving, and red dots mean vehicles have different speed values, corresponding stop-and-go waves.} 
\label{numfind}
\end{figure}

\section{The Case of Low Densities}\label{small}

The initial condition assumed for low densities is illustrated in Fig. \ref{Second}. While at large densities, vehicle $j=1$ had a {\it larger} distance than the other vehicles, it has now a {\it smaller} distance.
In the following, we will focus on vehicle $j=N$, of which we assume that it is located at $x=0$ and starts to accelerate at time $t=0$ with an initial speed $v_N(0)=0$, while the following vehicle $j=1$ is assumed to have the initial distance 
$d_1(0) = d_{\rm min}^{(0)} < d_0$. As the other vehicles have, by definition, the same initial distance $d_{(0)}$ from each other, the distances are related via the equation 
\begin{equation}
d_1(0) = d_{\rm min}^{(0)} = L - (N-1)d_{(0)} \, . 
\end{equation}
Equation (\ref{criamp}) determines again the perturbation size.
In the case of small densities $\overline{\rho}\le \rho_{\rm c2} = 1/d_0$, {\it the survival of a perturbation requires that there is at least one vehicle with a distance $d_{\rm min}^{(0)} < d_0$ 
for a long enough time period $t_*$ to force the successive vehicle to brake.} (If 
this vehicle would already start to accelerate before the follower reaches a distance $d_0$ to it, this would cause the perturbation to fade away.) Therefore, let us determine $t_*$ in the following.
\par\begin{figure}[htbp]
\begin{center}
\end{center}
\caption[]{Illustration of the initial distribution of vehicles assumed in the low-density case ($\overline{\rho} < 1/d_0$). Again, all vehicles are standing at time $t=0$, but vehicle $j=1$ has the initial distance $d_1(0) = d_{\rm min}^{(0)}=L-(N-1)d_{(0)}$, while all other vehicles have the initial distance $d_{(0)}>d_0$. The initial vehicle speeds were set to zero.}
\label{Second}
\end{figure}

\subsection{Derivation Focused on the Trajectory of One Vehicle}

According to Eqs. (\ref{cafo}) and (\ref{smpl}), the speed of vehicle $j=N$ evolves in time according to
$dv_N/dt = [v^0-v_N(t)]/\tau$, which implies
\begin{equation}
v_N(t) = \int\limits_0^t dt' \; \frac{dv_N(t')}{dt'} = v^0(1-\mbox{e}^{-t/\tau}) \approx v^0 \, , 
\end{equation}
and the related distance moved is
\begin{equation}
 d_N(t) = \int\limits_0^t dt' \; v(t') = v^0 t + v^0\tau(\mbox{e}^{-t/\tau} - 1) \approx v^0 \cdot (t-\tau) \, ,
 \label{e51}
\end{equation}
where the approximate equalities hold for $t \gg \tau$.
Let $d_{\rm min}^{(1)}$ denote the distance of the vehicles $j\in\{2, \dots, N-1\}$, 
after they were stopped by vehicle $j=1$ or a follower. 
Then, the distance moved by vehicle $j=N$ before it starts to decelerate 
at time $t=t_*$ (when it has reached a distance $d_0$ to its predecessor), is 
\begin{equation}
d_N(t_*) = L - d_{\rm min}^{(0)} - (N-2) d_{\rm min}^{(1)} - d_0 \, .
\label{e52}
\end{equation}
$d_{\rm min}^{(1)}$ denotes the minimum distance of a vehicle at the end of its first deceleration
maneuver, $d_{\rm min}^{(2)}$ after the second one, etc. 
Equations (\ref{e51}) and (\ref{e52}) imply the start of the deceleration maneuver at time $t=t_*^{(1)}$ with 
\begin{equation}
 t_*^{(1)} = \tau + \frac{L - d_{\rm min}^{(0)} - (N-2) d_{\rm min}^{(1)} - d_0}{v^0} \, .
 \end{equation}
In order for the perburbation to persist, this time period must be shorter than the time period $T_{(0)} + (N-2)T_{(1)}$,   
at which vehicle $j=1$ starts to accelerate, where $T_{(n)}$ is the time shift between the acceleration of two subsequent
vehicles standing at a distance $d_{\rm min}^{(n)}$. This implies the threshold condition
\begin{equation}
\tau + \frac{L - (N-1) d_{\rm min}^{(1)} - d_0}{v^0} =  T_{(0)} + (N-2)T_{(1)}
\label{this}
\end{equation}
for the survival of a perturbation. 
As the first deceleration maneuver extends over a time period $\Delta t = t_*^{(1)} - T_{(0)} - (N-2)T_{(1)}$, the evolution
$v_N(t) = v^0 \mbox{e}^{-(t-t_*^{(1)})/\tau}$ of the vehicle speed for $t_*^{(1)} < t \le t_*^{(1)}+\Delta t$ implies that the
minimum distance $d_{\rm min}^{(1)}$ afterwards is 
\begin{eqnarray}
 d_{\rm min}^{(1)} &=& d_0 - \!\!\!\! \int\limits_{t_*^{(1)}}^{t_*^{(1)}+\Delta t} \!\!\!\! dt' \; v_N(t') \nonumber \\
 &=& d_0 - v^0\tau(1 - \mbox{e}^{-\Delta t/\tau}) \ge d_0 - v^0 \tau \, .
 \label{demin} 
\end{eqnarray}
Moreover, let us assume a subsequent increase of the speed according to $v^0 (1-\mbox{e}^{-(t-t')/\tau})$, where $t'$ represents the starting time of the acceleration maneuver. Then, since the {\it following} vehicle starts to accelerate when that vehicle has reached a distance $d_0$, 
the time period $T_{(1)}$ between successive acceleration maneuvers is determined by the equation 
\begin{eqnarray}
 d_0 &=& d_{\rm min}^{(1)} + \int\limits_{t'}^{t'+T_{(1)}} dt \; v^0 ( 1 - \mbox{e}^{-(t-t')/\tau}) \nonumber \\
 &=& d_{\rm min}^{(1)} + v^0\Big[T_{(1)} - \tau ( 1 - \mbox{e}^{-T_{(1)}/\tau})\Big]  \, .
\label{secor}
\end{eqnarray}
Consider now that, for the critical amplitude, the conditions $d_{\rm min}^{(1)} = d_{\rm min}^{(0)}$ and $T_{(1)} = T_{(0)}$ of marginal stability must hold. Then, Eqs. (\ref{this}) and (\ref{secor}) imply 
\begin{eqnarray}
 (N-1)  T_{(0)} &=& (N-1) \left(\frac{d_0 - d_{\rm min}^{(0)} }{v^0} 
+ \tau ( 1 - \mbox{e}^{-T_{(0)}/\tau}) \right) \nonumber \\
&=& \frac{L - (N-1) d_{\rm min}^{(0)} - d_0}{v^0} \, .
\end{eqnarray}
Rearranging this and dividing the result by $N$, 
in the limit $N\gg 1$ and with $L/N = 1/\overline{\rho}$ we get
\begin{equation}
\frac{1}{\overline{\rho}} - d_0 \approx v^0  \tau ( 1 - \mbox{e}^{-T_{(0)}/\tau})  \, .
\label{apr1}
\end{equation}
Furthermore, with $d_{\rm min}^{(1)} = d_{\rm min}^{(0)}$ and $T_{(1)} = T_{(0)}$, Eq. (\ref{secor}) leads to 
\begin{equation}
d_{\rm min}^{(0)} - d_0 = v^0  \tau ( 1 - \mbox{e}^{-T_{(0)}/\tau}) - v^0T_{(0)} \approx v^0 \frac{(T_{(0)})^2}{2\tau} \, ,  
\label{apr2}
\end{equation}
where the last approximation is based on a Taylor expansion of second order. Therefore,
we have 
\begin{equation}
(v^0T_{(0)})^2 \approx 2\tau v^0(d_0 -d_{\rm min}^{(0)}) \label{apr3}
\end{equation}
(see Fig. \ref{see1}), while Eqs. (\ref{apr2}) and (\ref{apr1}) together imply
\begin{equation}
 v^0T_{(0)} \approx \frac{1}{\overline{\rho}} - d_{\rm min}^{(0)} \, . \label{apr4}
\end{equation}
As a consequence of Eqs. (\ref{apr3}) and (\ref{apr4}), we have
\begin{equation}
\left(\frac{1}{\overline{\rho}} - d_{\rm min}^{(0)}\right)^2 
\approx 2\tau v^0(d_0 -d_{\rm min}^{(0)}) \, . 
\label{criamp1}
\end{equation}
Solving this with respect to $d_{\rm min}^{(0)}$ finally gives
\begin{equation}
d_{\rm min}^{(0)} = - \left( \tau v^0 - \frac{1}{\overline{\rho}}\right) \pm \sqrt{ \Big( \tau v^0 - \frac{1}{\overline{\rho}}\Big)^2 - \Big( \frac{1}{\overline{\rho}^2}  - 2\tau v^0 d_0\Big)  } \, ,
\label{criamp2}
\end{equation}
but only the solution with the plus sign meets the requirement $d_{\rm min}^{(0)}> 0$ in the relevant density range (see Fig. \ref{see2}).
Considering Eqs. (\ref{criamp}) and (\ref{densconv}), we find the following relationship for the critical amplitude:
\begin{equation}
 \Delta \rho_{\rm cr} = \left| \frac{1}{d_{(0)}} - \frac{1}{d_1(0)} \right| \approx  
 \left| \overline{\rho} - \frac{1}{d_{\rm min}^{(0)}} \right| \, .
 \label{criamp3}
 \end{equation}
The critical amplitude ceases to be well-defined, when $d_{\rm min}^{(0)}$ assumes complex values, i.e. when the expression in Eq. (\ref{criamp2}) under the root becomes negative. The critical density corresponds $\rho_{\rm c1}$ to the density for which the root becomes zero, i.e. both solutions in Eq. (\ref{criamp2}) agree. This leads to 
\begin{equation}
\rho_{\rm c1} = \frac{1}{d_0 + \tau v^0/2} \, .
\label{rc1}
\end{equation}
With this, we have successfully estimated the critical density and the critical amplitudes, as Fig. \ref{numfind} shows. Notably, our results for the low-density regime deviate significantly from the lower limit
given in Eq. (\ref{nodiverge}). It is for the first time that the problem of determining critical perturbation amplitudes of a traffic model has been analytically solved.
\begin{figure}[htbp]
\begin{center}
\end{center}
\caption[]{(Color Online) Exact solution of Eq. (\ref{secor1}) as compared to its approximate solution (\ref{apr3}). The approximation works particularly well in the relevant range of $d_{\rm min}^{(0)}/d_0\approx 1$.}\label{see1}
\end{figure}
\begin{figure}[htbp]
\begin{center}
\end{center}
\caption[]{Both real-valued solutions of Eq. (\ref{criamp2}). It can be seen that the lower solution would predict negative values of $d_{\rm min}^{(0)}$ for most density values $\overline{\rho}$. Therefore, the upper solution must be chosen. Below $\overline{\rho} = \rho_{\rm c1}$, the solution of Eq. (\ref{criamp2}) has an imaginary part, which does not have a reasonable interpretation, here.}\label{see2}
\end{figure}

\subsection{Derivation Focused on A Leader and Its Follower}

While we have based our previous calculations on the study of vehicle $j=N$, we may also derive our results from the consideration of two successive vehicles instead. Analogously to Eq. (\ref{secor}), the time $T_{(0)}$ after which vehicle $j=1$ starts to {\it accelerate} is given by 
\begin{equation}
 d_{\rm min}^{(0)} + v^0\Big[T_{(0)} - \tau ( 1 - \mbox{e}^{-T_{(0)}/\tau})\Big] = d_0 \, .
\label{secor1}
\end{equation}
At that time, the following vehicle has the distance
\begin{equation}
d_{\rm min}^{(1)}  = \underbrace{d_{(0)} - v^0 t^{\prime\prime}}_{=d_0}  - v^0 \tau \big( 1 - \mbox{e}^{-(T_{(0)} - t^{\prime\prime})/\tau} \big) \, , 
\label{secor2}
\end{equation}
where
\begin{equation}
t^{\prime\prime} = \frac{d_{(0)} - d_0}{v^0} \approx \frac{1/\overline{\rho} - d_0}{v^0}
\label{tstr}
\end{equation}
is the time, at which vehicle $j=2$ starts to {\it decelerate}, while it moves at speed $v^0$ before. For the survival of the perturbation we have to demand $t^{\prime\prime} \ge 0$ and $d_{\rm min}^{(1)}\le d_{\rm min}^{(0)}$. In first-order Taylor approximation,
we have 
\begin{eqnarray}
d_0 - d_{\rm min}^{(1)} &=& v^0 \tau\big( 1 - \mbox{e}^{-(T_{(0)} - t^{\prime\prime})/\tau} \big) \nonumber \\
&\approx & v^0(T_{(0)} - t^{\prime\prime}) \approx  v^0T_{(0)} - \left(\frac{1}{\overline{\rho}} - d_0\right) \, ,\quad 
\end{eqnarray}
so that together with Eq. (\ref{secor2}) we get 
\begin{equation}
v^0 T_{(0)} = \frac{1}{\overline{\rho}} - d_{\rm min}^{(1)} \, . 
\label{equa}
\end{equation}
As pointed out before, for the critical amplitude we can presuppose marginal stability with $d_{\rm min}^{(1)} = d_{\rm min}^{(0)}$. Consequently, the relation $(v^0T_{(0)})^2 \approx 2\tau v^0(d_0 -d_{\rm min}^{(0)})$ results again as an approximate solution of (\ref{secor1}), see Fig. \ref{see3}.
\begin{figure}[htbp]
\begin{center}
\end{center}
\caption[]{Critical amplitude according to Eqs. (\ref{criamp3}) and (\ref{criamp2}). The curve is very well compatible with the numerical results, as shown in Fig. \ref{numfind}.}\label{see3}
\end{figure}

\section{Summary and Conclusions}\label{sumi}

Our findings can be summarized as follows: Based on the car-following model (\ref{cafo}) with the stepwise optimal velocity function (\ref{smpl}) and a simple type of initial perturbation, it is possible to derive critical amplitudes and critical densities. Linearly unstable behavior, where the breakdown of free traffic flow is caused by minor perturbations, is found only between the critical densities $\rho_{\rm c2}$ and $\rho_{\rm c3}$. In case of the stepwise optimal velocity function (\ref{smpl}), both critical densities agree, and we have $\rho_{\rm c2} = \rho_{\rm c3} = 1/d_0$. Below this density, the survival of a perturbation needs at least one vehicle with a distance $d_{\rm min}^{(0)} < d_0$ 
for a long enough time period to force the successive vehicle to brake. This requires densities between the critical densities $\rho_{\rm c1}$ and $\rho_{\rm c2}$, where $\rho_{\rm c1}$ is given by Eq. (\ref{rc1}) in good agreement with numerical results, see Fig. \ref{numfind}. Furthermore, the perturbations must have a size greater or equal to the critical perturbation amplitude $\Delta \rho_{\rm cr}(\overline{\rho})$. Equation (\ref{criamp3}) with (\ref{criamp2}) represent the related results of our approximate analytical calculations. 
\par
If the average density $\overline{\rho}$ exceeds the value $1/d_0$, 
an initial perturbation will fade away, if the maximum vehicle distance does not allow drivers to reach a sufficiently high speed in the acceleration process to reach a vehicle distance equal to or smaller than $d_{(0)}$ after the successive braking maneuver. With Eq. (\ref{rc4}), we have found an analytical formula for the critical density $\rho_{\rm c4}$, above which perturbations will necessarily decay. However, for average densities between $\rho_{\rm c3}$ and $\rho_{\rm c4}$, perturbations larger than the critical amplitude given by Eq. (\ref{biig}) will grow and form moving jams of the kind of ``anticlusters'', while smaller perturbations will fade away, giving rise to homegonous, congested traffic (and in the special case of the stepwise optimal velocity function assumed here, even to standing traffic). Note that,  for other specifications of the optimal velocity function, homogeneous congested traffic flows will usually be {\it finite}. Furthermore, it should not be forgotten that the critical perturbation amplitude may depend on the shape of the initial perburbation. 
\par
Our analytical findings are summarized by Fig. \ref{numfind}. 
Note that the self-organized flow-density relation (the ``jam line'') differs significantly from the ``fundamental diagram'' resulting from the optimal velocity function in the case of stationary and homogeneous traffic flows, which, however, may be unstable. An interesting observation for the stepwise specification of the optimal velocity function is that large perturbations at high vehicle densities can reach {\it greater} average flows than {\it small} perturbations, while the situation at low densities is characterized by a capacity drop, see Eq. (\ref{DroP}).

\begin{acknowledgement}
The authors would like to thank Peter Felten for the preparation of illustrations \ref{First} and \ref{Second}.\\
{\it Author Contributions:} MM produced the other figures and performed the computer simulations, while DH did the analytical calculations.
\end{acknowledgement}

\appendix
\section{Derivation of the Characteristic Constants of the Optimal Velocity Model}

For the stepwise specification (\ref{smpl}), the characteristic constants of the optimal velocity model (\ref{cafo})
can be analytically calculated (see Ref. \cite{Exact}). In the following, we summarize the calculations in our notation, here. For this, let us assume a number of jammed vehicles with velocity zero and distance smaller than $d_0$. $\rho_{\rm jam}$ shall denote the jam density. 
\par
If we assume that a car starts to accelerate out of the jam only when its leading car has approximately reached its desired velocity $v^0$, and starts to decelerate when its predecessor has almost stopped in the traffic jam,
we have the acceleration equation $dv_j/dt \approx [v^0 - v_j(t)]/\tau > 0$ for $d_j>d_0$ and $dv_j/dt \approx [0-v_j(t)]/\tau < 0$
for $d\le d_0$. Consequently, we find 
\begin{equation}
 v_j(t) \approx \left\{
 \begin{array}{ll}
 v^0 \big(1 - \mbox{e}^{-(t-t_0)/\tau}\big) & \mbox{if } d_j(t) > d_0 \, , \\
 v^0 \mbox{e}^{-(t-t_2)/\tau} & \mbox{otherwise.}
 \end{array}\right.
\end{equation}
$t_0$ is the time point when the acceleration of vehicle $j$ starts with with $d_j(t_0) = d_0$ and $dd_j(t_0)/dt > 0$, while $t_2$ is the successive time point with $d(t_2)=d_0$ and $dd_j(t_2)/dt < 0$, when the deceleration starts. 
\par\begin{figure}[htbp]
  \begin{center}
\end{center}
\caption[]{Illustration of the approximate periodic change of distance and speed (``hysteresis loop'') for fully developed
traffic jams  (solid lines with arrows), after Ref. \cite{Exact}. The dashed line indicates the underlying optimal velocity function
$v_{\rm o}(d)$, which is very different.\label{ILL}}
\end{figure}
We may distinguish four different phases:
\begin{itemize}
\item Phase 1 is characterized by $dv_{j-1}(t)/dt > 0$ and $dv_{j}(t)/dt > 0$, i.e. vehicle $j$ and its leader $j-1$ both accelerate. According to the travelling wave concept, the trajectory of vehicle $j$ is exactly identical with the one of its predecessor, but shifted by some time period $T$, which corresponds to the delay between the acceleration of two successive vehicles out of the traffic jam, i.e. 
\begin{equation}
v_j(t) = v_{j-1}(t-T) \, . 
\end{equation}
The speed $c$ of jam resolution is then given by the distance $-1/\rho_{\rm jam}$ between two jammed vehicles, divided by this time period $T$:
\begin{equation}
 c = - \frac{1}{\rho_{\rm jam}T} \, .
 \label{ccalc}
\end{equation}
The negative sign is a consequence of the fact that vehicle $j$ stands upstream of vehicle $j-1$, but accelerates later, so that the downstream jam front travels opposite to the direction of motion. 
\par
Considering the delay $T$, we get 
\begin{eqnarray}
\frac{dd_j(t)}{dt} &=& v_{j-1}(t) - v_j(t) \nonumber \\
&=& v_j(t-T) - v_j(t) \, , 
\end{eqnarray}
which by integration over time and with $d_j(t_0) = d_0$ results in 
\begin{eqnarray}
 d_j(t) &=& d_0 + \big( 1 - \mbox{e}^{-T/\tau}\big) v^0 \tau \Big( 1 - \mbox{e}^{-(t-t_0)/\tau} \Big) \nonumber \\
 &=& d_0 + \big( 1 - \mbox{e}^{-T/\tau}\big) \tau v_j(t) \, .
\end{eqnarray}
This implies a linear increase of distance with speed (see Fig. \ref{ILL}).
Consequently, the desired velocity $v_j(t) = v^0$ is reached at the distance
\begin{equation}
\frac{1}{\rho_{\rm out}} = d_0 +  \tau v^0\big( 1 - \mbox{e}^{-T/\tau}\big) \, ,
\label{quout1}
\end{equation}
which defines the density $\rho_{\rm out}$ related with the characteristic outflow $Q_{\rm out} = \rho_{\rm out} v^0$ from congested traffic. 
\item Phase 2 is characterized by $dv_{j-1}(t)/dt < 0$ and $dv_{j}(t)/dt > 0$, i.e. vehicle $j-1$ already decelerates, while vehicle $j$ still accelerates. This phase is assumed to start at time $t_1 > t_0$, and we have $dd_j(t)/dt = v^0\mbox{e}^{-(t-t_1)/\tau}
- v^0 (1-\mbox{e}^{-(t-t_0)/\tau})<0$. 
Therefore, we get
\begin{eqnarray}
d_j(t) &=& d_j(t_1)  -  v^0\tau (\mbox{e}^{-(t-t_1)/\tau} - 1)- v^0\cdot(t-t_1) \nonumber \\[1mm]
 &-& v^0 \tau ( \mbox{e}^{-(t-t_0)/\tau} - \mbox{e}^{-(t_1-t_0)/\tau} )\nonumber \\
 &\approx & d_j(t_1) - v^0(t-t_1) + v^0 \tau \Big(1-\mbox{e}^{-(t-t_1)/\tau} \Big) \, ,\qquad 
\end{eqnarray}
where the approximation assumes $t - t_0 \ge t_1-t_0 \gg \tau$.
That is, the distance $d_j(t)$ goes down, while the speed $v_j(t)= v^0(1-\mbox{e}^{-(t-t_0)/\tau})$ is approximately $v^0$.
The distance $d_j(t)$ becomes $d_0$ at time $t_2$ with 
\begin{equation}
 (t_2 - t_1) = \frac{d_j(t_1)-d_0}{v^0} + \tau \, . 
\end{equation}
\item Phase 3 starts at time $t_2 > t_1$ and is characterized by $dv_{j-1}(t)/dt < 0$ and $dv_{j}(t)/dt < 0$, i.e. vehicle $j-1$ decelerates, and vehicle $j$ does the same with a time delay of $T$. One can say that phase 3 is the inverse process of phase 1, and we get
\begin{eqnarray}
 d_j(t) &=& d_0 - v^0\big( 1 - \mbox{e}^{-T/\tau}\big) \tau \Big( 1 - \mbox{e}^{-(t-t_2)/\tau} \Big) \nonumber \\
 &=& d_0 - \big( 1 - \mbox{e}^{-T/\tau}\big) \tau \big[ v^0 - v_j(t) \big] \, .
\end{eqnarray}
Accordingly, the distance is monotonously decreasing with time. The minimum distance is reached at time $t_3$ for the jam density $\rho_{\rm jam}$, which is defined by
\begin{equation}
 \frac{1}{\rho_{\rm jam}} = d_0 -  \tau v^0 \big( 1 - \mbox{e}^{-T/\tau}\big)  \, .
\label{jamdef1}
\end{equation}
Together with Eq. (\ref{quout1}) we find
\begin{equation}
 \frac{1}{\rho_{\rm out}} - \frac{1}{\rho_{\rm jam}} = 2\tau v^0  \big( 1 - \mbox{e}^{-T/\tau}\big) \, .
\end{equation}
\item Phase 4 starts at time $t_3$ and is characterized by $dv_{j-1}(t)/dt > 0$ and $dv_{j}(t)/dt < 0$, i.e. vehicle $j$ still decelerates, while its leader $j-1$ already accelerates. Phase 4 is the inverse process of phase 2, and we have $dd_j/dt = v^0(1-\mbox{e}^{-(t-t_3)/\tau}) - v^0 \mbox{e}^{-(t-t_2)/\tau}$. Therefore, we find
\begin{eqnarray}
 d_j(t) &=& d_j(t_3) + v^0\cdot(t-t_3) +  v^0\tau (\mbox{e}^{-(t-t_3)/\tau} - 1) \nonumber \\[1mm]
 &+& v^0 \tau ( \mbox{e}^{-(t-t_2)/\tau} - \mbox{e}^{-(t_3-t_2)/\tau} )\nonumber \\
 &\approx & \frac{1}{\rho_{\rm jam}} + v^0(t-t_3) - v^0 \tau \Big(1-\mbox{e}^{-(t-t_3)/\tau} \Big) \, ,\qquad
\end{eqnarray}
where the approximation assumes $t-t_3\ge t_3 -t_2 \gg \tau$.
Therefore, the distance grows in time, while the velocity $v_j(t) = v^0\mbox{e}^{-(t-t_2)/\tau}$ of vehicle $j$ is approximately zero. At time $t_4$, the distance $d_j(t)$ becomes $d_0$. The difference $t_4-t_3$ determines the time delay $T$ between two successive acceleration maneuvers of cars leaving the traffic jam. Inserting the definition (\ref{jamdef1}) 
for $\rho_{\rm jam}$, we obtain
\begin{eqnarray}
 d_0 &=&  \frac{1}{\rho_{\rm jam}} + v^0 T - \tau v^0\Big(1-\mbox{e}^{-T/\tau} \Big) \nonumber \\
&=&  d_0 + v^0 T - 2\tau v^0\Big( 1 - \mbox{e}^{-T/\tau}\Big) 
\end{eqnarray}
and the implicit relationship for the time shift
\begin{equation}
 T=  2 \tau\Big( 1 - \mbox{e}^{-T/\tau}\Big) \, .
\label{Tedef1}
\end{equation}
Together with Eq. (\ref{quout1}), this yields
\begin{equation}
\frac{1}{\rho_{\rm out}} = d_0 + \frac{v^0T}{2} \, . 
\label{twentyonea}
\end{equation}
\end{itemize}
Therefore, according to Eqs. (\ref{jamdef1}), (\ref{quout1}), and (\ref{Tedef1}), it is possible to express the characteristic constants $T$, $c$, $\rho_{\rm jam}$, and $\rho_{\rm out}$ through the model parameters $d_0$, $v^0$, and $\tau$ only, without any dependence on the initial condition. Note that all the characteristic parameters are no model parameters. They are rather a result of the self-organization of characteristic jam fronts (see Fig. \ref{ILL}). 
\end{document}